\documentstyle[12pt,aaspp4]{article}

\begin{document}

%
\def\n{\footnotemark}
\def\IUE{{\it IUE}}
\def\HST{{\it HST}}
\def\ISO{{\it ISO}}
\def\deg{$^{\rm o}$}
\def\degC{$^{\rm o}$C}
\def\arcsec{\ifmmode '' \else $''$\fi}
\def\arcmin{$'$}
\def\arcsecpoint{\ifmmode ''\!. \else $''\!.$\fi}
\def\arcminpoint{$'\!.$}
\def\kms{\ifmmode {\rm km\ s}^{-1} \else km s$^{-1}$\fi}
\def\Msun{\ifmmode {\rm M}_{\odot} \else M$_{\odot}$\fi}
\def\Lsun{\ifmmode {\rm L}_{\odot} \else L$_{\odot}$\fi}
\def\Zsun{\ifmmode {\rm Z}_{\odot} \else Z$_{\odot}$\fi}
\def\ergsAcm{ergs\,s$^{-1}$\,cm$^{-2}$\,\AA$^{-1}$}
\def\ergscm2{ergs\,s$^{-1}$\,cm$^{-2}$}
\def\qo{\ifmmode q_{\rm o} \else $q_{\rm o}$\fi}
\def\Ho{\ifmmode H_{\rm o} \else $H_{\rm o}$\fi}
\def\ho{\ifmmode h_{\rm o} \else $h_{\rm o}$\fi}
\def\ltsim{\raisebox{-.5ex}{$\;\stackrel{<}{\sim}\;$}}
\def\gtsim{\raisebox{-.5ex}{$\;\stackrel{>}{\sim}\;$}}
\def\vFWHM{\ifmmode v_{\mbox{\tiny FWHM}} \else
            $v_{\mbox{\tiny FWHM}}$\fi}
\def\CCF{\ifmmode F_{\it CCF} \else $F_{\it CCF}$\fi}
\def\ACF{\ifmmode F_{\it ACF} \else $F_{\it ACF}$\fi}
\def\Halpha{\ifmmode {\rm H}\alpha \else H$\alpha$\fi}
\def\Hbeta{\ifmmode {\rm H}\beta \else H$\beta$\fi}
\def\Hgamma{\ifmmode {\rm H}\gamma \else H$\gamma$\fi}
\def\Hdelta{\ifmmode {\rm H}\delta \else H$\delta$\fi}
\def\Lya{\ifmmode {\rm Ly}\alpha \else Ly$\alpha$\fi}
\def\Lyb{\ifmmode {\rm Ly}\beta \else Ly$\beta$\fi}
\def\Lyg{\ifmmode {\rm Ly}\beta \else Ly$\gamma$\fi}
\def\hi{H\,{\sc i}}
\def\hii{H\,{\sc ii}}
\def\hei{He\,{\sc i}}
\def\heii{He\,{\sc ii}}
\def\ci{C\,{\sc i}}
\def\cii{C\,{\sc ii}}
\def\ciii{\ifmmode {\rm C}\,{\sc iii} \else C\,{\sc iii}\fi}
\def\civ{\ifmmode {\rm C}\,{\sc iv} \else C\,{\sc iv}\fi}
\def\ni{N\,{\sc i}}
\def\nii{N\,{\sc ii}}
\def\niii{N\,{\sc iii}}
\def\niv{N\,{\sc iv}}
\def\nv{N\,{\sc v}}
\def\oi{O\,{\sc i}}
\def\oii{O\,{\sc ii}}
\def\oiii{O\,{\sc iii}}
\def\o5007{[O\,{\sc iii}]\,$\lambda5007$}
\def\oiv{O\,{\sc iv}}
\def\ov{O\,{\sc v}}
\def\ovi{O\,{\sc vi}}
\def\neiii{Ne\,{\sc iii}}
\def\nev{Ne\,{\sc v}}
\def\neviii{Ne\,{\sc viii}}
\def\mgi{Mg\,{\sc i}}
\def\mgii{Mg\,{\sc ii}}
\def\mgx{Mg\,{\sc x}}
\def\siIV{Si\,{\sc iv}}
\def\siIII{Si\,{\sc iii}}
\def\siII{Si\,{\sc ii}}
\def\si{S\,{\sc i}}
\def\sii{S\,{\sc ii}}
\def\siii{S\,{\sc iii}}
\def\siv{S\,{\sc iv}}
\def\sv{S\,{\sc v}}
\def\svi{S\,{\sc vi}}
\def\caii{Ca\,{\sc ii}}
\def\feii{Fe\,{\sc ii}}
\def\feiii{Fe\,{\sc iii}}
\def\alii{Al\,{\sc ii}}
\def\aliii{Al\,{\sc iii}}
\def\piv{P\,{\sc iv}}
\def\pv{P\,{\sc v}}
\def\cliv{Cl\,{\sc iv}}
\def\clv{Cl\,{\sc v}}
\def\nai{Na\,{\sc i}}
\def\o{\o}
%

\title{Do the Broad Emission Line Clouds See the Same Continuum that We See?}

\author{Kirk Korista \& Gary Ferland}
\affil{Department of Physics \& Astronomy, University of Kentucky,
Lexington, KY 40506}
\author{Jack Baldwin}
\affil{Cerro Tololo Interamerican Observatory\altaffilmark{1}, Casilla
603, La Serena, Chile}

\altaffiltext{1}{Operated by the Association of Universities for
Research in Astronomy Inc.\ (AURA) under cooperative agreement with the
National Science Foundation}

\begin{abstract}

Recent observations of quasars, Mrk~335 and the \HST\/ quasar composite
spectrum, have indicated that many of them have remarkably soft
ionizing continua ($f_{\nu} \propto \nu^{-2}$, 13.6~eV -- 100~eV). We
point out that the number of $h\nu > 54.4~eV$ photons is insufficient
to create the observed strengths of the \heii\/ emission lines. While
the numbers of photons which energize \civ\/ $\lambda$1549 and \ovi\/
$\lambda$1034 are sufficient, even the most efficiently emitting clouds
for these two lines must each cover at least 20\% -- 40\% of the
source. If the typical quasar ionizing continuum is indeed this soft,
then we must conclude that the broad emission line clouds must see a
very different (harder) continuum than we see. The other viable
possibility is that the UV -- EUV SED is double-peaked, with the second
peaking near 54~eV, its Wien tail the observed soft X-ray excess.

\end{abstract}

%
%
%
\section{Introduction}

Since their recognition in 1963 as the most luminous entities in the
universe, the continuum spectral energy distributions (SEDs) of quasars
and active galactic nuclei have been studied extensively as a means of
understanding these energetic objects (e.g., Malkan \& Sargent 1982;
Sanders et al.\ 1989; Elvis et al.\ 1995). Of secondary importance is
that this continuum ionizes and excites the gas that emits the observed
broad emission lines (Davidson 1977). Important to both is knowledge of
the strength and shape of the ionizing continuum (13.6~eV -- 1~KeV).

Determining the SED and number of ionizing photons has been difficult,
since most of the ionizing continuum lies in the unobservable portions
of the spectrum. This continuum had to be inferred based upon the
strengths of observed emission lines, such as \heii\/ $\lambda$1640
(Mathews \& Ferland 1987), and others (Binette, Courvoisier, \&
Robinson 1988; Krolik \& Kallman 1988; Binette et al.\ 1989; Clavel \&
Santos-Lle\'{o} 1990; Zheng 1991; Gondhalekar 1992; Zheng, Fang, \&
Binette 1992). However, with the advent of high quality ground-based
and space-based observations, the ``unobservable gap'' has begun to
narrow substantially from the low and high energy ends of the ionizing
spectrum. Based upon some of these recent contemporaneous broad band
continuum observations, we will consider the continuum -- broad
emission line energy budgets of quasars, and discuss whether or not
these {\em observed} continuum SEDs might differ significantly from
those {\em incident} upon the broad emission line clouds.

Contemporaneous ground-based, \IUE\/, {\em HUT}, and {\em BBXRT}
observations, led Zheng et al.\ (1995) to propose that a comptonized
accretion disk produces the continuum SED for Mrk~335. Their proposed
continuum is one which, beginning near the rest-frame Lyman limit
(912~\AA\/), falls as $f_{\nu} \propto \nu^{-2}$, before meeting the
X-ray power law (see also Mannheim et al.\ 1995).

More recently, Zheng et al.\ (1997) presented a composite \HST\/ quasar
spectrum. The quasar redshifts of $z \gtsim 1$ allowed them to observe
into the rest frame far-UV. The analysis of Zheng et al.\ showed that
an SED very similar to that inferred for Mrk~335 was consistent with
their composite quasar spectrum, corrected for Galactic reddening and
intergalactic absorption. Based upon the results of Zheng et al.\ and
their quasar soft X-ray composite spectrum, Laor et al.\ (1997) also
suggested the absence of an EUV bump.

This very soft UV-bump may apply to high redshift quasars as well.
Multi-wavelength observations by Bechtold et al.\ (1994), Elvis et
al.\ (1995), and Kuhn et al.\ (1995) presented similar evidence for
soft Lyman continua in these objects after correcting for Galactic
reddening and intergalactic absorption. They inferred a natural
connection between their data near the Lyman limit and the X-ray
observations, the inferred slope roughly that as the comptonized
accretion disk spectrum model of Zheng et al.\ (1995).

Such a soft ionizing continuum is surprising, especially in Mrk~335
that historically is known to be a relatively strong emitter of
\heii\/. Zheng et al.\ (1995) found
$W_{\lambda}$(\heii\/~$\lambda\/1640) \approx 11.5$~\AA\/, while from
the \IUE\/ archives Koratkar (1990) found an average equivalent with of
about 13~\AA\/ in this line.  Over a 6 year period 1989--1990 Kassebaum
et al.\ (1997) found $W_{\lambda}$(\heii\/~$\lambda\/4686) \approx
22$~\AA\/ (see also Shuder 1981; Boroson \& Green 1992).  In this
paper, we will show that due to the paucity of 54.4~eV photons, this
SED is unlikely to account for the observed equivalent width of \heii\/
$\lambda$1640 in Mrk~335 or in the composite \HST\/ quasar spectrum.
Thus if the observed SEDs have been inferred correctly, {\em the broad
emission line clouds must see a different, harder continuum than we
observe}. We present the calculations in $\S$~2. A discussion of this
conundrum and other possible explanations and a summary follow in
$\S$~3 and $\S$~4.

\section{The Maximum \heii\/ $\lambda$1640 Equivalent Width}

Our goal is to search for those cloud parameters (gas density, ionizing
flux) which maximize the emission of \heii\/ $\lambda$1640, and compare
the energy emitted to that observed.  Of course, one could produce an
approximate theoretical limit to the \heii\/ equivalent width based
upon the number of photons at energies greater than 54.4~eV, and
assuming Case~B emissivity (MacAlpine 1981; MacAlpine et al.\ 1985;
Mathews \& Ferland 1987).  However, collisional excitation as well as
optical depth effects will modify this result, and so we have chosen to
run numerical simulations to identify the maximum theoretical limit to
the \heii\/ $\lambda$1640 equivalent width for a given incident SED.
While we rely on the numerical results, we note that the predicted
emission was essentially Case~B except at high densities where the
predicted flux was $\sim 2\times$ larger and at high incident flux
where thermalization at large optical depth depressed the emission
below Case~B.

Using the photoionization code {\sc Cloudy} (c90.03; Ferland 1996) we
computed a grid of broad emission line spectra in the gas density --
hydrogen ionizing flux plane (see Korista et al.\ 1997). Using a
piece-wise power law fit to the spectrum of Mrk~335 inferred by Zheng
et al.\ (1995), spectra from 841 clouds were computed. We assumed an
$\alpha_{ox} \approx 1.46$, approximately matching the observations of
Mrk~335 reported by Zheng et al.\ (1995).  This value of $\alpha_{ox}$
is rather soft for a Seyfert~1, but typical for moderate luminosity
QSOs such as those in the \HST\/ quasar composite (Wilkes et al.\ 1994;
Laor et al.\ 1997). The gas density and ionizing flux spanned 7 orders
of magnitude along each axis, and we assumed a column density of
$10^{23}$~cm$^{-2}$ and solar abundances. A larger column density will
not increase the maximum equivalent width of \heii\/ $\lambda$1640, and
higher metal abundances will reduce its equivalent width due to the
increased metal opacity (Ferland et al.\ 1996). In Figure~1 we plot in
the gas density -- hydrogen ionizing flux plane contours of four
emission line equivalent widths, referenced to the incident continuum
at 1215~\AA\/, as in Baldwin et al.\ (1995) and Korista et
al.\ (1997).  The cloud covering fraction ($f_c$) is assumed full in
each case. A comparison of these line equivalent widths with those from
the latter two references demonstrates the diminished emission line
equivalent widths resulting from the very soft spectrum employed here.
Note that contour ratios in Figure~1 are equivalent to line flux
ratios, since the line equivalent widths are all referenced to the same
continuum point at 1215~\AA\/.

As indicated in Figure~1, these calculations show that this soft
continuum generates a {\bf maximum} equivalent width (with $f_c = 1$)
in \heii\/ $\lambda$1640 of $\approx 13$~\AA\/, as measured at
1640~\AA\/. From Figure~1, the cloud that emits \heii\/ $\lambda$1640
optimally has $\log n_H \approx 12.75$ and $\log \Phi(H) \approx 21$
corresponding to $\log U(H) \approx -2.25$. Here, $\Phi(H)$ is the
hydrogen ionizing photon flux and $U(H) \equiv \Phi(H)/(n_Hc)$. The
observed equivalent widths in this line in Mrk~335 and the composite
\HST\/ quasar spectrum are $\approx 11.5$~\AA\/ and $\approx
4.5$~\AA\/, respectively. The Mrk~335 measurement contains a minority
narrow line contribution, but this matters little since sufficient
numbers of 54.4~eV photons still need to emerge from the nucleus to
ionize the narrow line clouds. The case of Mrk~335 would require nearly
full global coverage of the source by these {\em optimally emitting}
\heii\/ clouds. However, if the column densities of these particular
clouds were comparable to those expected for typical broad line clouds
($10^{22-24}$~cm$^{-2}$), this shell would be opaque between 13.6~eV
and a few KeV, while such absorption is never seen.  Additionally, the
presence of strong \ovi\/ $\lambda$1034 and significant
intercombination line strengths in the spectrum of Mrk~335 and the
quasar composite spectrum require populations of clouds with other
parameters. Thus the emitting clouds of the BLR cannot all be those
that emit \heii\/ optimally (see Figure~1), and an ensemble of BLR
clouds exposed to this soft continuum is more likely (e.g., Baldwin et
al.\ 1995).  Integrating over an ensemble of clouds leads to predicted
\heii\/ $\lambda$1640 equivalent widths of $\approx$~3--4~\AA\/ for
$f_c = 0.5$, much lower than the observed equivalent width in Mrk~335.
Thus, the inferred SED at 54.4~eV of Mrk~335 is too weak by factors of
several.

For the \HST\/ quasar composite we can consider statistical arguments
concerning the cloud covering fraction. Smith et al. (1981), Kinney et
al.\ (1985), Antonucci et al.\ (1989), Koratkar et al.\ (1992) and
others searched for and did not find intrinsic Lyman limit absorption
in the spectra of quasars, arguing for cloud covering fractions of
$\sim$~10\%. For this covering fraction and the predicted
$W_{\lambda}$(\heii\/~$\lambda\/1640) \approx$ 6 -- 8~\AA\/ $\times f_c
= 0.6 - 0.8$~\AA\/ the \HST\/ quasar composite spectrum's SED at
54.4~eV is at least 5 times too weak; covering fractions of 56\% --
75\% are required.

These calculations assumed thermal intrinsic line widths. If
significant turbulence or streaming is present within the emitting gas,
then the line optical depths will diminish (perhaps along preferred
directions).  However, we find that the introduction of non-thermal
line widths does not change the maximum equivalent width of \heii\/
$\lambda$1640, but merely shifts this maximum to clouds exposed to
higher ionizing fluxes.

We conclude that this soft SED cannot account for the observed
strengths of the \heii\/ emission lines in quasars and AGNs, at least
under the usual assumption that the cloud covering fraction is less
than 1. 

\section{Discussion}

\subsection{Incident EUV Flux Brighter Than Observed?}

Either the inferred continua are incorrect {\em or else the clouds see
a brighter EUV bump than we do}. The latter is possible if the clouds
are located in a flattened geometry, for example, and we view the
quasar preferentially from above in the presence of an intrinsically
anisotropic continuum. The predicted SEDs of simple accretion disk
models do predict the disk to become harder at larger inclinations due
to relativistic effects (Sun \& Malkan 1989).  Scenarios such as this
have been discussed before for broad line clouds, on theoretical bases
(e.g., Netzer 1987), as well as observational ones (Ferland et
al.\ 1996). An ionizing photon deficit became apparent more than a
decade ago in the observations of extended narrow line regions of
Seyfert galaxies (Neugebauer et al.\ 1980; Wilson et al.\ 1988; Penston
et al.\ 1990); a comprehensive review of this problem is given by
Binette, Fosbury, \& Parker (1993). However, this is the first time
that observations have provided such a straightforward and stringent
test for the broad emission line clouds. If true this has important
consequences to our understanding of quasars.

\subsection{Is the UV-EUV Bump``Double-Peaked''?}

It is to be emphasized that the spectrum of an AGN or quasar has never
been observed at 54~eV in the rest-frame. Thus, a second possibility is
that the SED at energies of $\sim$~40~eV -- 300~eV is rather
fortuitously dominated by a {\em second, powerful, EUV bump}, peaking
near 54~eV. In order to power the HeII line in Mrk~335, it would have
to contain energy comparable to the ``classical'' UV-bump. There is
evidence for such power in the EUV in at least one object. Mrk~478 is
1--3 times brighter in $\nu\/L_{\nu}$ at $\sim$~120~eV than it is at
1300~\AA\/ (Gondhalekar et al.\ 1994; Marshall et al.\ 1996). At higher
energies, $\sim$~200~eV -- 400~eV, a soft X-ray excess has been
observed in many AGN and quasars (e.g., Saxton et al.\ 1993). Perhaps
this excess is not the tail of the classical UV-bump (or associated
accretion disk), but the Wien tail of a separate EUV-bump. This
EUV-bump would have to lie just beyond the soft Lyman continuum which
extends to at least 30~eV in the composite \HST\/ quasar spectrum, and
its Wien tail would be the observed soft X-ray excess. A single
blackbody with $T_{BB} \sim 2 \times 10^5$~K might accomplish this. The
origin of such a component is unclear, but could be the result of
reprocessing (Czerny \& \.{Z}ycki 1994).

\subsection{Another Energy Source?}

Another possibility is that another source of energy is available to
the clouds other than photoionization from the central source, such as
has been suggested for some of the the \feii\/ emission in quasars
(Collin-Souffrin 1986; Joly 1987). This does not seem likely for
\heii\/, however, based upon the results of the line-continuum
reverberation campaigns for AGNs (e.g., Korista et al.\ 1995),
including Mrk~335 (Kassebaum et al.\ 1997), that show a strong
correlation between the strength of \heii\/ and the continuum.

\subsection{Low Column Density He~II Emitting Clouds?}

In $\S$~2 we concluded that even if there exists a special type of
\heii\/ emitting cloud population that fully covers the continuum
source, it would be opaque to ionizing photons for standard broad
emission line cloud column densities ($N(H) \approx
10^{22-24}$~cm$^{-2}$). Our calculations show that the clouds which
emit \heii\/ $\lambda$1640 optimally (Figure~1) barely reproduce the
observed equivalent width in Mrk~335 with a minimum column density
$N(H) \approx 10^{20.3}$ and full source coverage. Such clouds are
optically thick between 54.4~eV and $\sim$~200~eV, but optically thin
at all other energies. With the exception of those lines that are
emitted in the He$^{++}$ zone, such a He$^+$ Lyman continuum filter
would have relatively minor effects on the emitted spectrum of clouds
that lie outside the filter. \civ\/ $\lambda$1549 would emit at roughly
2/3 strength in clouds that lie outside the filter. Thus in this way,
well-layered clouds with chosen parameters would allow the integrated
covering fraction of the BLR to be greater than 1, and the observed
\heii\/ strength could be reconciled with the soft ionizing SED.

There are at least two problems with this scenario, however. First, it
is improbable that this soft ionizing SED can simultaneously account
for the observed \heii\/ $\lambda$1640 and \ovi\/ $\lambda$1034
equivalent widths. The observed \ovi\/ (broad line) equivalent widths
of 23~\AA\/ in Mrk~335 and 16~\AA\/ in the quasar composite require the
clouds that emit \ovi\/ {\em most efficiently} (Figure~1) to cover 30\%
of the source in Mrk~335 and typically 20\% of the source in the quasar
composite spectrum. These same clouds must have larger ionization
parameters and column densities ($\sim 10^{22-23}$~cm$^{-2}$) than the
optimal \heii\/ emitting clouds, and thus be completely opaque to the
He$^+$ Lyman continuum out to a few KeV. Since no \ovi\/ is emitted in
the optimal \heii\/ emitting clouds (Figure~1), the \ovi\/ emitters
must lie mainly {\em inside} the \heii\/ emitting shell (some of the
\ovi\/ emission could originate outside the \heii\/ emitting shell, but
at $\ltsim 1/3$ efficiency due to the shell's opacity). In the case of
Mrk~335, 30\% of the He$^+$ Lyman continuum photons that were required
to illuminate the \heii\/ emitting shell would be absorbed by the
\ovi\/ emitting clouds inside the shell. The equivalent width of
\heii\/ $\lambda$1640 emitted in the \ovi\/ clouds amounts to $0.3
\times 8$~\AA\/, not quite enough to account for the deficit of \heii\/
emission in the shell due to the opacity of the \ovi\/ clouds.

A related problem is the observed \civ\/ equivalent widths. As a
bolometer \civ\/ $\lambda$1549 is less sensitive to the gas
abundances.  Its observed equivalent widths of 50~\AA\/ and 69~\AA\/ in
Mrk~335 and the quasar composite spectrum require the most optimally
emitting \civ\/ clouds (Figure~1) to cover 28\% and 38\% of the source,
respectively. The required source coverages become larger in the
presence of less efficient \civ\/ emitting clouds and larger still in
the presence of an interior \heii\/ emitting shell. However, in
contrast to \heii\/ $\lambda$1640, significant non-thermal gas motions
would increase the peak equivalent widths of \civ\/ and \ovi\/,
reducing somewhat their required cloud covering fractions.

A final problem with this scenario is that if the bulk of \heii\/ is
emitted in this special set of low column density clouds, their
response to continuum variations would likely be negative, and
certainly not strongly positive, contrary to what has been observed in
Mrk~335 (Kassebaum 1997) and other Seyfert~1 galaxies (Clavel et
al.\ 1991).

\subsection{Possible Pitfalls in the Interpretations of the
Observations}

It is possible that the far-UV continuum of the \HST\/ Quasar composite
was not correctly inferred. Significant corrections for intervening
absorption had to be made to the data. Zheng et al.\ (1997) claim an
uncertainty in their 350~\AA\/ -- 1050~\AA\/ slope of $\pm$0.15;
however, even if this were an underestimate, they find consistency with
previous measurements of far-UV slopes in other quasars.

It should also be noted that the composite \HST\/ quasar spectrum is
composed of a varying number of constituents at different wavelengths.
Specifically, while most objects in the composite contributed to
\heii\/ $\lambda$1640, a diminishing number contributed to wavelengths
shorter than about 1000~\AA\/.  Thus, determining how representative
the composite spectrum is of the individual quasars is not
straightforward. In particular, it is unclear from a composite spectrum
how the \heii\/ strengths in {\em individual} objects correlate with
the far-UV SEDs.  Although in the case of Mrk~335 there was minimal
far-UV spectral coverage, the contemporaneous multi-wavelength nature
of these observations mitigates the ``composite'' problem and it
demonstrates a larger deficit of 54.4~eV photons.

\section{Summary}

In summary, recent observations of quasars have indicated many of them
to have remarkably soft ionizing continua, such that the number of
54.4~eV photons is insufficient to create the observed strengths of the
\heii\/ emission lines. While the numbers of photons which energize
\civ\/ $\lambda$1549 and \ovi\/ $\lambda$1034 are sufficient, even the
most efficiently emitting clouds for these two lines must each cover at
least 20\% -- 40\% of the source. If the soft ionizing continua
conclusion holds up under further scrutiny and are common, then we
consider the most likely conclusion to be that the broad emission line
clouds must see a very different (harder) continuum than we see. The
other viable option would seem to be that the UV -- EUV SED is
double-peaked, with the second peaking near 54~eV, its Wien tail the
observed soft X-ray excess.

\acknowledgements This work was supported by the NSF (AST 93-19034),
NASA (NAGW-3315, NAG-3223), and STScI grant GO-2306. We thank an
anonymous referee for his or her constructive comments.
%
%
%

%
\newpage
%
%
%
\begin{figure}
\plotfiddle{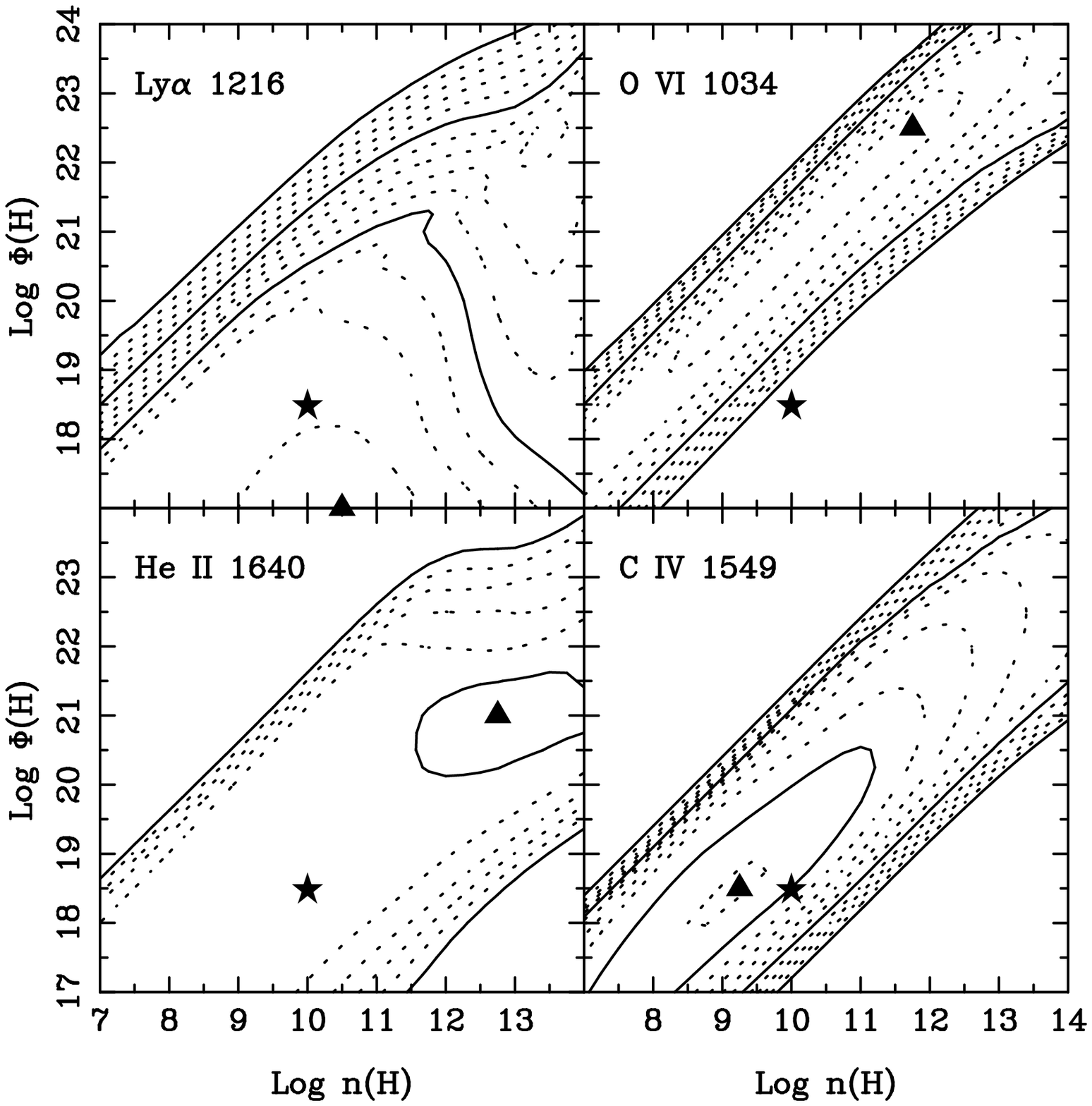}{5.0in}{0}{70}{70}{-205}{-110}
\caption{Contours of $\log W_{\lambda}$ for four emission lines for
the soft SED grid are shown as a function of the hydrogen density and
flux of hydrogen ionizing photons. Line strengths are expressed as
logarithmic equivalent widths referenced to the incident continuum at
1215~\AA\/ for full source coverage. The smallest decade contoured is
1~\AA\/, each solid line is 1 dex, and dotted lines represent 0.2 dex
steps. In each case the peak in the equivalent width distribution lies
beneath the center of a solid triangle. The contours generally decrease
monotonically from the peak to the 1~\AA\/ contour.  The solid star is
a reference point marking the ``standard BLR'' parameters discussed by
Davidson and Netzer (1979).}
\end{figure}
\end{document}